\documentclass[a4paper,twoside,12pt]{article}
\input{obsjkt.sty}

\newcommand{\Msun}{\ensuremath{~{\rm M}_\odot}}                   
\newcommand{\Rsun}{\ensuremath{~{\rm R}_\odot}}                   
\newcommand{\rhosun}{\ensuremath{~\rho_\odot}}                    
\newcommand{\Teff}{\ensuremath{T_{\rm eff}}}                      
\newcommand{\TeffA}{\ensuremath{T_{\rm eff,A}}}                   
\newcommand{\TeffB}{\ensuremath{T_{\rm eff,B}}}                   
\newcommand{\Vsys}{\ensuremath{V_\gamma}}                         
\newcommand{\EBV}{\ensuremath{E(B\!-\!V)}}                        
\newcommand{\Grp}{\ensuremath{G_{\rm RP}}}                        
\newcommand{\degr}{\ensuremath{^\circ}}                           
\renewcommand{\kms}{~km~s$^{-1}$}                                 
\renewcommand{\cd}{~d$^{-1}$}                                     
\newcommand{\chir}{\ensuremath{\chi_\nu^{\,2}}}                   
\newcommand{\mc}[1]{\multicolumn{2}{c}{#1}}

\newcommand{\kepler}{\textit{Kepler}}
\newcommand{\hip}{\textit{Hipparcos}}
\newcommand{\gaia}{\textit{Gaia}}
\newcommand{\targ}{OO~Peg}
\newcommand{\targfull}{OO~Pegasi}

\newcommand{\Msunnom}{\hbox{$\mathcal{M}^{\rm N}_\odot$}}
\newcommand{\Rsunnom}{\hbox{$\mathcal{R}^{\rm N}_\odot$}}
\newcommand{\Lsunnom}{\hbox{$\mathcal{L}^{\rm N}_\odot$}}

\usepackage{rotating}

\begin{document} 

\OBSheader{Rediscussion of eclipsing binaries: \targ}{J.\ Southworth}{2024 June}

\OBStitle{Rediscussion of eclipsing binaries. Paper XVIII. \\ The F-type system OO Pegasi}

\OBSauth{John Southworth}

\OBSinstone{Astrophysics Group, Keele University, Staffordshire, ST5 5BG, UK}


\OBSabstract{\targ\ is a detached eclipsing binary system containing two late-A-type stars in a circular orbit with a period of 2.985~d. Using published spectroscopic results and a light curve from the Transiting Exoplanet Survey Satellite (TESS) we determine their masses to be $1.69 \pm 0.09$ and $1.74 \pm 0.06$\Msun\ and their radii to be $2.12 \pm 0.03$ and $1.91 \pm 0.03$\Rsun. The TESS data are of high quality, but discrepancies in the radial velocities from two sources prevent a precise mass measurement. The primary star is definitively hotter, larger and more luminous than its companion, but its mass is lower (albeit to a significance of only 1.1$\sigma$). Using published apparent magnitudes and temperatures, we find a distance of $238.8 \pm 6.1$~pc, in agreement with the \gaia\ DR3 parallax. Although both components are in the $\delta$~Scuti instability strip, we find no evidence of pulsations. More extensive spectroscopy is needed to improve our understanding of the system.}


\section*{Introduction}

In this series of papers \cite{Me20obs} we have been systematically reanalysing known detached eclipsing binaries (dEBs) in order to determine their physical properties to high precision. The main improvements versus previous work stem from the availability of high-quality light curves from space missions such as \kepler\ \cite{Borucki16rpph} and TESS (Transiting Exoplanet Survey Satellite \cite{Ricker+15jatis}) -- see ref.~\cite{Me21univ} for a review. 

This work is important because dEBs are our primary source of direct measurements of the basic properties (mass and radius)  of normal stars \cite{Andersen91aarv,Torres++10aarv}. They are widely used to calibrate physical processes included in theoretical models of stellar evolution \cite{LastennetVallsgabaud02aa,DelburgoAllende18mn}, such as atomic diffusion \cite{HiglWeiss17aa}, convective core overshooting \cite{ClaretTorres18apj} and the size of the core \cite{Tkachenko+20aa}. A high precision in the measurements of the stellar properties is vital for reliable results \cite{Valle+18aa} and can approach 0.2\% precision in mass and radius in the best cases \cite{Maxted+20mn}.


\section*{\targfull}

\begin{table}[t]
\caption{\em Basic information on \targfull. \label{tab:info}}
\centering
\begin{tabular}{lll}
{\em Property}                            & {\em Value}                 & {\em Reference}                   \\[3pt]
Right ascension (J2000)                   & 21:41:37.70                 & \cite{Gaia21aa}                   \\
Declination (J2000)                       & +14:39:30.8                 & \cite{Gaia21aa}                   \\
Henry Draper designation                  & HD 206417                   & \cite{CannonPickering24anhar}     \\
\textit{Hipparcos} designation            & HIP 107099                  & \cite{Hipparcos97}                \\
\textit{Gaia} DR3 designation             & 1770729907069675392         & \cite{Gaia21aa}                   \\
\textit{Gaia} DR3 parallax                & $4.2534 \pm 0.0245$ mas     & \cite{Gaia21aa}                   \\          
TESS\ Input Catalog designation           & TIC 314847177               & \cite{Stassun+19aj}               \\
$U$ magnitude                             & $8.650 \pm 0.010$           & \cite{Oja87aas}                   \\          
$B$ magnitude                             & $8.635 \pm 0.021$           & \cite{Hog+00aa}                   \\          
$V$ magnitude                             & $8.354 \pm 0.018$           & \cite{Hog+00aa}                   \\          
$J$ magnitude                             & $7.676 \pm 0.023$           & \cite{Cutri+03book}               \\
$H$ magnitude                             & $7.633 \pm 0.027$           & \cite{Cutri+03book}               \\
$K_s$ magnitude                           & $7.555 \pm 0.018$           & \cite{Cutri+03book}               \\
Spectral type                             & A7~V + A8~V                 & \cite{Cakirli15newa}              \\[3pt]
\end{tabular}
\end{table}



In this work we present an analysis of \targfull\ (Table~\ref{tab:info}) based on published spectroscopy and new space-based photometry. The eclipsing nature of \targ\ was found using data from the \hip\ satellite \cite{Hipparcos97,Kazarovets+99ibvs}. A first detailed analysis was presented by Munari et al.\ \cite{Munari+01aa} (hereafter M01) with the aim of assessing the expected quality of results from the then-forthcoming \gaia\ mission; of the three systems in that paper V505~Per and V570~Per have already been revisited by the current author using the new TESS data \cite{Me21obs5,Me23obs4}. M01 used only \hip\ photometry and radial velocity (RV) measurements from ground-based spectroscopy in the 850--875~nm region, to represent the type of observations that \gaia\ was expected to obtain. Due to these limitations they were only able to obtain masses to 2\% and radii to 4\% precision. 

A subsequent analysis of \targ\ was presented by \c{C}ak{\i}rl{\i} \cite{Cakirli15newa} (hereafter C15) who added new spectroscopic RV measurements and a more extensive light curve from the All Sky Automated Survey (ASAS \cite{Pojmanski97aca}) to determine the properties of the components more precisely. C15 also measured the atmospheric parameters of the components, their projected rotational velocities, and the reddening and distance of the system. He searched for but found no evidence of pulsations in the light curves from \hip\ and ASAS, despite both components being in the $\delta$~Scuti instability strip \citep{Murphy+19mn}.

The apparent mangitudes in Table~\ref{tab:info} come from a variety of sources. The $U$ magnitude is from Oja \cite{Oja87aas} and relies on just two observations so may not reflect the brightness of the system outside eclipse. This number is not used in our analysis, but the consistency between the distances measured in the various passbands (see below) suggest it does represent an out-of-eclipse measurement. The $BV$ magnitudes are from the Tycho experiment \cite{Hog+00aa} on the \hip\ satellite and each comprise the average of 55 measurements well-distributed in orbital phase. The $JHK_s$ magnitudes are from 2MASS \cite{Cutri+03book} and were obtained at a single epoch corresponding to orbital phase 0.615, which is not within an eclipse.



\section*{Photometric observations}

\begin{figure}[t] \centering \includegraphics[width=\textwidth]{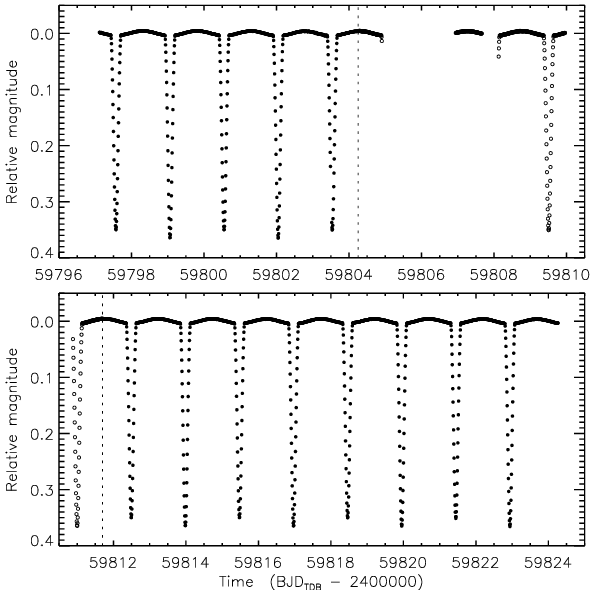} \\
\caption{\label{fig:time} TESS\ short-cadence SAP photometry of \targ. The flux 
measurements have been converted to magnitude units then rectified to zero magnitude 
by subtraction of the median. The data rejected from the analysis are shown using open
circles, and the corresponding cutoff times indicated with vertical dashed lines.} \end{figure}

%

\targ\ has been observed just once by TESS, in sector 55, beginning on 2022/08/05 and concluding on 2022/09/01. A second set of observations is scheduled for sector 82 and will occur in 2024 August if the spacecraft remains healthy. The observations from sector 55 were obtained with a cadence of 600~s, which is lower than desired and decreases the information content of the data.

The available light curve from TESS was downloaded from the NASA Mikulski Archive for Space Telescopes (MAST\footnote{\texttt{https://mast.stsci.edu/portal/Mashup/Clients/Mast/Portal.html}}) using the {\sc lightkurve} package \cite{Lightkurve18}. We used the simple aperture photometry (SAP) data from the TESS-SPOC data reduction \cite{Jenkins+16spie}. A quality flag of ``hard'' yielded a total of 3412 datapoints (Fig.~\ref{fig:time}). We rejected the data in the time interval BJD$_{\rm TDB}$ 2459804.25 to 2459811.70 to avoid a stretch of data with the eclipses either partially covered or not observed at all, leaving 2831 datapoints for further analysis. These were normalised using {\sc lightkurve}, converted to differential magnitude, and the median magnitude of the sector subtracted.

A query of the \gaia\ DR3 database\footnote{\texttt{https://vizier.cds.unistra.fr/viz-bin/VizieR-3?-source=I/355/gaiadr3}} returned a total of 56 objects within 2~arcmin of \targ. Of these, the brightest is fainter by 4.23~mag in the $G$ band and 3.98~mag in the \Grp\ band. A small amount of third light is therefore expected to contaminate the TESS light curve of \targ, at the level of approximately 1\%.


\section*{Light curve analysis}

\begin{figure}[t] \centering \includegraphics[width=\textwidth]{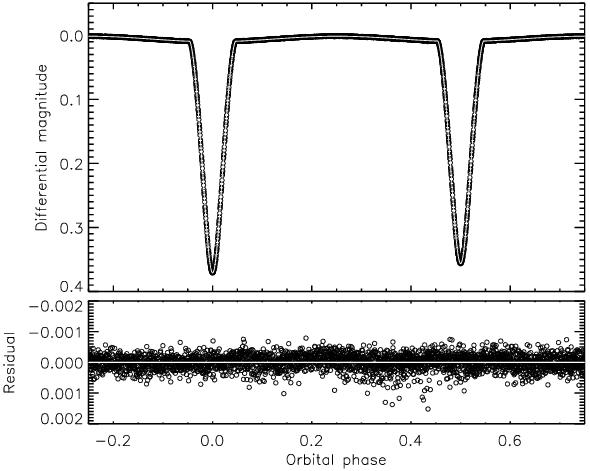} \\
\caption{\label{fig:phase} The 600-s cadence TESS light curves of \targ\ (filled 
circles) and its best fit from {\sc jktebop} (white-on-black line) versus orbital 
phase. The residuals are shown on an enlarged scale in the lower panel.} \end{figure}

The components of \targ\ are small compared to their orbital separation, so the system is suitable for analysis with the {\sc jktebop}\footnote{\texttt{http://www.astro.keele.ac.uk/jkt/codes/jktebop.html}} code \cite{Me++04mn2,Me13aa}, for which we used version 43. We defined the primary eclipse to be the deeper of the two eclipses, star~A to be the component eclipsed at primary eclipse, star~B to be its companion, and the primary eclipse to occur at orbital phase zero. In the case of \targ\ star~A is both hotter and larger than star~B, by small but significant amounts.

The {\sc jktebop} fitted parameters included the fractional radii of the stars ($r_{\rm A}$ and $r_{\rm B}$), expressed as their sum ($r_{\rm A}+r_{\rm B}$) and ratio ($k = {r_{\rm B}}/{r_{\rm A}}$), the central surface brightness ratio ($J$), orbital inclination ($i$), orbital period ($P$), and a reference time of primary minimum ($T_0$). A circular orbit was assumed, after confirming that allowing for orbital eccentricity has a negligible effect on the values of the fitted parameters. 

Initial attempts to fit for third light returned values that were very small and slightly less than zero. An experiment with it fixed to a value of 2\%, to account for the nearby stars discussed above, yielded a solution with significantly larger residuals and a noticably poorer fit to the eclipses. We therefore fixed third light to zero. 

Limb darkening was included using the power-2 law \cite{Hestroffer97aa,Maxted18aa,Me23obs2} defined according to
\begin{equation}
\frac{F(\mu)}{F(1)} = 1 - c(1-\mu^\alpha)
\end{equation}
where $\mu = \cos\gamma$, $\gamma$ is the angle between the observer's line of sight and the surface normal, $F(\mu)$ is the surface brightness at position $\mu$ on the stellar disc, $c$ is the linear coefficient and $\alpha$ is the nonlinear coefficient. As the two stars are very similar we assumed their limb-darkening behaviours to be identical. Initial fits showed that we were able to fit for one but not both of the limb darkening coefficients, so we fitted for $c$ and left $\alpha$ fixed at a theoretical value \cite{ClaretSouthworth22aa,ClaretSouthworth23aa}.

The relatively low 600~s sampling rate of the TESS data was accounted for by numerically integrating the model to match \cite{Me11mn}. In effect we calculated the model at five points, each spaced by 120~s, and averaged the results before comparing to an observed datapoint. We found that this had a negligible effect on the results, but continued to do so as the increase in computation time was not a problem. The coefficients of two quadratic functions, one for each half of the TESS sector, were also included to precisely normalise the light curve to zero differential magnitude.

We found no evidence for changes in the orbital period for \targ, in agreement with the results of C15. We therefore included the observed time of primary minimum from the \hip\ light curve calculated by M01 ($2448499.1545  \pm 0.0020$) to help constrain the orbital ephemeris more precisely. This step lowered the uncertainty in $P$ by approximately a factor of 3.

\begin{table} \centering
\caption{\em \label{tab:jktebop} Parameters of \targ, with their 1$\sigma$ uncertainties, 
measured from the TESS sector 55 light curves using the {\sc jktebop} code.}
\setlength{\tabcolsep}{4pt}
\begin{tabular}{lcc}
{\em Parameter}                           &              {\em Value}            \\[3pt]
{\it Fitted parameters:} \\                                                   
Primary eclipse time (BJD$_{\rm TDB}$)    & $ 2459813.984151   \pm 0.000007   $ \\
Orbital period (d)                        & $       2.98465593 \pm 0.00000049 $ \\
Orbital inclination (\degr)               & $      83.629      \pm  0.013     $ \\
Sum of the fractional radii               & $       0.30576    \pm  0.00020   $ \\
Ratio of the radii                        & $       0.8983     \pm  0.0038    $ \\
Central surface brightness ratio          & $       0.96661    \pm  0.00016   $ \\
LD coefficient $c$                        & $       0.709      \pm  0.014     $ \\
LD coefficient $\alpha$                   &            0.431 (fixed)            \\
{\it Derived parameters:} \\                                                   
Fractional radius of star~A               & $       0.16107    \pm  0.00028   $ \\
Fractional radius of star~B               & $       0.14489    \pm  0.00039   $ \\
Light ratio $\ell_{\rm B}/\ell_{\rm A}$   & $       0.7794     \pm  0.0065    $ \\[3pt]
\end{tabular}
\end{table}


The resulting best fit is shown in Fig.~\ref{fig:phase} and the parameters are given in Table~\ref{tab:jktebop}. Uncertainties in the fitted parameters were calculated using both Monte Carlo and residual-permutation simulations \cite{Me++04mn,Me08mn}, and the larger of the two options chosen for each parameter. The two error estimation algorithms were in good agreement for all parameters, as expected because there is no obvious systematic noise present in the data. The uncertainties in the all-important fractional radii are encouragingly low at 0.17\% and 0.27\%, despite the relatively poor sampling rate of the TESS photometry.


\section*{Radial velocity analysis}

M01 published a set of 21 RVs for each component of \targ, which are tabulated in the paper. The spectra on which they were based were deliberately obtained at quasi-random times in order to simulate a dataset that might be expected from \gaia. As a result, two spectra are too blended to give precise RVs and there is only one spectrum near second quadrature. We reanalysed the RVs from M01 to confirm their results, and followed these authors in omitting the RVs from the two most blended spectra. 

To fit the RVs we used {\sc jktebop} and the orbital ephemeris from Table~\ref{tab:jktebop}. The fitted parameters were the velocity amplitudes of the two stars, $K_{\rm A}$ and $K_{\rm B}$, and the systemic velocity (\Vsys) of the two stars. We also allowed for a change in $T_0$ to insure against ephemeris drift or period changes, but all solutions were consistent with the ephemeris given in Table~\ref{tab:jktebop}. Separate fits were obtained with \Vsys\ either assumed to be the same for the two stars or allowed to be different. The errorbars of the RVs for each star were scaled to give a reduced $\chi^2$ of $\chir=1.0$ versus the best fit. Parameter uncertainties were calculated using the Monte Carlo procedure \cite{Me21obs5}. The results are given in Table~\ref{tab:orbits}.

It was immediately clear that our star~A is the \emph{secondary} component for M01, something that can happen easily when the primary and secondary eclipses are of similar depth and the photometric data are quite scattered. We accounted for this in our analysis. We note that this is also apparent in C15 (see his fig.~4) but not commented on by that author. Our $K_{\rm A}$ agrees with M01, but our $K_{\rm B}$ and \Vsys\ do not agree within the uncertainties. We also notice that there is a counterintuitive result that the r.m.s.\ residuals are lower for star~A when the \Vsys\ values of the two stars are required to be the same -- this occurs because of the rescaling of the RV uncertainties combined with the RV measurements having a range of uncertainties.

\begin{sidewaystable} \centering
\caption{\em \label{tab:orbits} Spectroscopic orbits for \targ\ from the literature and from the reanalysis 
of the RVs in the current work. $K_{\rm A}$ and $K_{\rm B}$ values were not given by M01, so we have 
calculated the values that would reproduce their mass measurements. All quantities are in km~s$^{-1}$.}
\begin{tabular}{lccccccc}
{\em Source}  & $K_{\rm A}$~ & $K_{\rm B}$~ & ${\Vsys}$~ & ${\Vsys}_{\rm ,A}$~ & ${\Vsys}_{\rm ,B}$~ & \mc{r.m.s.\ residual} \\
              &              &              &            &                     &                     & star~A & star~B \\[8pt]
M01	(our calculation)				&      112.0      &      110.1      & $8.47 \pm 0.46$ &                &                &     &     \\
C15									& $111   \pm 2  $ & $114   \pm 3  $ & $6.7  \pm 0.1 $ &                &                &     &     \\[5pt]
M01 (our fit, \Vsys\ same)			& $112.1 \pm 1.6$ & $107.6 \pm 1.0$ & $10.8 \pm 0.7$  &                &                & 5.1 & 6.7 \\
M01 (our fit, \Vsys\ different)		& $112.4 \pm 1.7$ & $107.6 \pm 1.0$ &                 & $11.4 \pm 1.2$ & $10.5 \pm 0.8$ & 5.4 & 6.6 \\[5pt]
C15 (our fit, \Vsys\ same)			& $112.9 \pm 1.6$ & $111.8 \pm 1.2$ & $5.5  \pm 0.9$  &                &                & 5.1 & 3.2 \\
C15 (our fit, \Vsys\ different)		& $112.3 \pm 1.5$ & $111.5 \pm 1.3$ &                 & $ 7.6 \pm 1.2$ & $ 3.9 \pm 1.2$ & 4.2 & 3.1 \\[5pt]
Combined (\Vsys\ same)				& $112.7 \pm 1.1$ & $109.1 \pm 0.9$ & $0.2  \pm 0.7$  &                &                & 5.1 & 5.9 \\
Combined (\Vsys\ different)			& $112.5 \pm 1.0$ & $109.0 \pm 0.9$ &                 & $ 0.3 \pm 0.8$ & $-0.4 \pm 0.7$ & 5.1 & 5.6 \\[5pt]
Adopted values						& $112.6 \pm 1.2$ & $109.1 \pm 2.8$ &                 &                &                &     &     \\
\end{tabular}
\end{sidewaystable}

C15 obtained 15 spectra of \targ\ from which 15 RVs were obtained for star~A and 14 for star~B. C15 fitted spectroscopic orbits to the spectra from his own observations together with those from M01. We first modelled the RVs from C15 separately. Four of the RVs for star~A and three for star~B are close to conjunction so suffer from blending and contribute little to pinning down $K_{\rm A}$ and $K_{\rm B}$, so we ran solutions with these omitted. The results were similar to those for all RVs, and we adopted these as our standard datasets for the RVs from C15.

\begin{figure}[t] \centering \includegraphics[width=\textwidth]{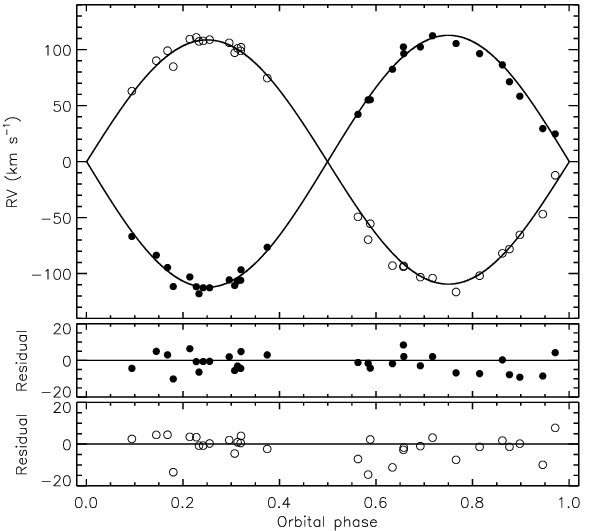} \\
\caption{\label{fig:rv} RVs of \targ\ from M01 and C15 (filled circles for star~A 
and open circles for star~B), compared to the best fit from {\sc jktebop} (solid 
lines) with a separate \Vsys\ value for each star. The residuals are given in the 
lower panels separately for the two components.} \end{figure}

Table~\ref{tab:orbits} shows that there are significant discrepancies between the solutions of different RV datasets, both calculated in this work and versus the literature. We also consistently find that $K_{\rm A}$ is larger than $K_{\rm B}$, thus star~A is less massive than star~B (although the difference is of similar size to the uncertainties). Some of these discrepancies are driven by small-number statistics, and some are likely due to differences in \Vsys\ from the differing RV measurement processes used by M01 and C15. The biggest discrepancy is in the $K_{\rm B}$ values from the two sources of RVs, which differ by 4\kms.

We made the choice to fit the RVs from M01 and C15 simultaneously, both with the combined and independent \Vsys\ values (Fig.~\ref{fig:rv}). In each case we scaled the errorbars of the individual datasets to give $\chir=1$ and subtracted the best-fitting \Vsys\ before fitting the combined data. Our results are in between those for the two RV sources separately, as expected. The $K_{\rm A}$ values are consistent over all solutions so we adopt a value of $112.6 \pm 1.2$\kms. The errorbar is the quadrature addition of the 1.1\kms\ uncertainty in Table~\ref{tab:orbits} and the 0.5\kms\ which is the largest difference between the adopted $K_{\rm A}$ and the other fitted values. For $K_{\rm B}$ we adopt a value of $109.1 \pm 2.8$\kms, where the uncertainty is the quadrature addition of 0.9\kms\ and 2.7\kms\ following the same argument. This $K_{\rm B}$ is unfortunately rather uncertain, which prevents the measurement of the masses of the stars to high precision.


\section*{Physical properties and distance to \targ}

\begin{table} \centering
\caption{\em Physical properties of \targ\ defined using the nominal solar units given by 
IAU 2015 Resolution B3 (ref.\ \cite{Prsa+16aj}). The \Teff\ values are from C15. \label{tab:absdim}}
\begin{tabular}{lr@{\,$\pm$\,}lr@{\,$\pm$\,}l}
{\em Parameter}        & \multicolumn{2}{c}{\em Star A} & \multicolumn{2}{c}{\em Star B}    \\[3pt]
Mass ratio   $M_{\rm B}/M_{\rm A}$          & \multicolumn{4}{c}{$1.032 \pm 0.029$}         \\
Semimajor axis of relative orbit (\Rsunnom) & \multicolumn{4}{c}{$13.16 \pm 0.18$}          \\
Mass (\Msunnom)                             &  1.689  & 0.088       &  1.744  & 0.058       \\
Radius (\Rsunnom)                           &  2.120  & 0.029       &  1.907  & 0.027       \\
Surface gravity ($\log$[cgs])               &  4.013  & 0.011       &  4.119  & 0.005       \\
Density ($\!\!$\rhosun)                     &  0.1774 & 0.0027      &  0.2515 & 0.0040      \\
Synchronous rotational velocity ($\!\!$\kms)& 35.93   & 0.50        & 32.32   & 0.45        \\
Effective temperature (K)                   &  7850   & 350         &  7600   & 450         \\
Luminosity $\log(L/\Lsunnom)$               &   1.19  & 0.08        &  1.04   & 0.10        \\
$M_{\rm bol}$ (mag)                         &   1.77  & 0.20        &  2.14   & 0.26        \\
Interstellar reddening \EBV\ (mag)			& \multicolumn{4}{c}{$0.09 \pm 0.02$}			\\
Distance (pc)                               & \multicolumn{4}{c}{$238.8 \pm 6.1$}           \\[3pt]
\end{tabular}
\end{table}


Using the photometric and spectroscopic results from Tables \ref{tab:info}, \ref{tab:jktebop} and \ref{tab:orbits}, we have determined the physical properties of the \targ\ system using the {\sc jktabsdim} code \cite{Me++05aa}. The results are given in Table~\ref{tab:absdim} and show that the masses are measured to 5.2\% (star~A) and 3.3\% (star~B), and the radii to 1.4\% (both stars). This is not the desired 2\% precision \cite{Andersen91aarv,Me15debcat} due to the uncertainty in the value of $K_{\rm B}$. The mass measurements agree well with those of M01 but not C15; the radius measurements disagree with both. In particular, the $R_{\rm B}$ value from M01 ($1.37 \pm 0.05$\Rsun) is extremely low. Our results are based on a careful analysis of the available RVs and much higher-quality light curves from TESS, so should be preferred to previous values.

M01 determined the effective temperatures (\Teff s) of the stars from a comparison between the observed and synthetic spectra, finding $\TeffA = 8770 \pm 150$~K and $\TeffB = 8683 \pm 180$~K. The ratio of these values agrees well with the surface brightness ratio measured from the light curve (Table~\ref{tab:jktebop}). Using these \Teff s and the apparent magnitudes of the system (Table~\ref{tab:info}), we determined the distance to \targ\ using the $K$-band surface brightness method \cite{Me++05aa} and calibrations from Kervella et al.\ \cite{Kervella+04aa}. The interstellar reddening was determined by requiring the $UBV$ and $JHK$ distances to agree, via manual iteration, resulting in $\EBV = 0.21 \pm 0.03$~mag and a distance of $245.2 \pm 4.9$~pc. This reddening is rather larger than expected -- the {\sc stilism}\footnote{\texttt{https://stilism.obspm.fr}} online tool \cite{Lallement+14aa,Lallement+18aa} gives a value of $0.037 \pm 0.018$~mag -- and the distance is also $2\sigma$ beyond the value \gaia\ DR3 \cite{Gaia21aa} value of $234.1 \pm 1.3$~pc.

C15 determined rather smaller temperatures of $\TeffA = 7850 \pm 350$~K and $\TeffB = 7600 \pm 450$~K, via comparison with reference-star spectra. Using these values instead of the ones from M01, we obtain $\EBV = 0.09 \pm 0.02$~mag and a distance of $238.8 \pm 6.1$~pc. This \EBV\ is in much better agreement with the {\sc stilism} value, and the distance is also consistent with the \gaia\ DR3 parallax at the 0.8$\sigma$ level. We therefore adopt these \Teff s and \EBV\ as our final values in Table~\ref{tab:absdim}. Supporting evidence for these lower temperatures are the catalogue \Teff s of $7476 \pm 149$~K given in v8 of the TESS Input Catalog \cite{Stassun+19aj} and $7347 \pm 17$~K from the \gaia\ DR3 APSIS pipeline \cite{Creevey+23aa,Fouesneau+23aa}. Both catalogues treat point sources as single stars, but in the case of \targ\ this is a reasonable approximation due to the similarity of the two components.

C15 measured $\EBV = 0.29 \pm 0.01$ from the strength of the interstellar Na D lines; such a large reddening is highly inconsistent with our results and would require the stars to have \Teff s in the region of 10,000~K for the distances measured in the optical to match those measured in the IR. 


\section*{Summary and final points}

\targ\ is a dEB containing two component with late-A spectral types, on a circular orbit with a period of 2.98~d, whose eclipsing nature was discovered thanks to the \hip\ satellite. We have presented a reanalysis of the system based on a space-based light curve from the TESS mission and published spectroscopic parameters. The TESS light curve is of high quality and allows the fractional radii of the stars to be determined to 0.2\% precision; star~A is clearly larger, hotter and more luminous than its companion. However, our reanalysis of published RVs from two sources yields both a disagreement in the value of $K_{\rm B}$ and the measurement of a lower mass for star~A than star~B. This discrepancy would be problematic for stellar evolutionary theory, but is thankfully not significant due to the uncertainty in the measured masses. Using published apparent magnitudes of the system and \Teff\ values of the stars, we have determined a distance to the system in agreement with the \gaia\ DR3 parallax.

Both components of \targ\ are in a region of the luminosity versus \Teff\ diagram where a high fraction of stars show $\delta$~Scuti pulsations \cite{Murphy+19mn}, but are not known to pulsate. We therefore calculated a Fourier transform of the residuals of the {\sc jktebop} fit using version 1.2.0 of the {\sc period04} code \cite{LenzBreger05coast}. No significant periodicity was found up to the Nyquist frequency of 72\cd, with a noise level of approximately 0.01~mmag from 1\cd\ to the Nyquist frequency. Lower-amplitude pulsations may be present but would require significantly more photometry to measure.

We made a brief comparison of the masses, radii and \Teff s of the stars to the {\sc parsec} 1.2S theoretical stellar evolutionary models \cite{Bressan+12mn,Chen+14mn}. Agreement was found for a solar chemical composition and an age of $1.0 \pm 0.3$~Gyr, which supports the lower \Teff\ values found by C15 versus those obtained by M01.

The quality of our results has been limited by the imprecision of the spectroscopic parameters measured for the system: both the RVs and the \Teff s are quite uncertain. Conversely, the TESS data allow high-quality measurements of its photometric parameters. Further work should therefore concentrate on performing a more extensive spectroscopic analysis of \targ. Forthcoming data releases from the \gaia\ satellite may well help.


\section*{Acknowledgements}

We thank the anonymous referee for a helpful report.
This paper includes data collected by the TESS\ mission and obtained from the MAST data archive at the Space Telescope Science Institute (STScI). Funding for the TESS\ mission is provided by the NASA's Science Mission Directorate. STScI is operated by the Association of Universities for Research in Astronomy, Inc., under NASA contract NAS 5–26555.
This work has made use of data from the European Space Agency (ESA) mission {\it Gaia}\footnote{\texttt{https://www.cosmos.esa.int/gaia}}, processed by the {\it Gaia} Data Processing and Analysis Consortium (DPAC\footnote{\texttt{https://www.cosmos.esa.int/web/gaia/dpac/consortium}}). Funding for the DPAC has been provided by national institutions, in particular the institutions participating in the {\it Gaia} Multilateral Agreement.
The following resources were used in the course of this work: the NASA Astrophysics Data System; the SIMBAD database operated at CDS, Strasbourg, France; and the ar$\chi$iv scientific paper preprint service operated by Cornell University.



\end{document}